\documentclass[12pt]{article}

\usepackage{amsmath,amssymb,amsbsy,amsfonts,latexsym,graphicx}
\usepackage{hyperref}
\usepackage{color, array, subfigure}
\usepackage{cite}

\allowdisplaybreaks

\evensidemargin \oddsidemargin

\begin{document}

%\newcommand{\story}{\vspace{5mm} \noindent $\spadesuit$ }

%%%%%%%%%%%%%%%%%%%%%%%%%%% TITLE PAGE %%%%%%%%%%%%%%%%%%%%%%%%%%%%%%%
\begin{titlepage}

%-------------------- footnote symbol in title page -----------------
\renewcommand{\thefootnote}{\fnsymbol{footnote}}

%----------------------- preprint number & date ---------------------

%\begin{flushright}
%???-??????
%\end{flushright}

%---------------------------- Title ---------------------------------
\vspace{15mm}
\baselineskip 9mm
\begin{center}
  {\Large \bf 
{  
End of the World Perspective to BCFT}
 }
\end{center}

%--------------------- Authors and Addresses ------------------------
\baselineskip 6mm
\vspace{10mm}
\begin{center}
Kyung Kiu Kim$^{1,a}$, Sejin Kim$^{1,b}$, Jung Hun Lee$^{1,c}$, Chanyong Park$^{2,d}$ and Yunseok Seo$^{1,e}$
 \\[10mm] 
  $^1${\sl College of General Education, Kookmin University, Seoul 02707, Korea}
   \\[3mm]
    $^2${\sl School of Physics and Chemistry, Gwangju Institute of Science and Technology,\\Gwangju 61005, Korea}
    \\[3mm]
  {\tt  ${}^a$kimkyungkiu@kookmin.ac.kr, ${}^b$sejin817@kookmin.ac.kr, ${}^c$junghun.lee@kookmin.ac.kr, ${}^d$cyong21@gist.ac.kr, ${}^e$yseo@kookmin.ac.kr
  }
\end{center}

\thispagestyle{empty}

%-------------------------- abstract --------------------------------
%\vfill
\vspace{1cm}
\begin{center}
{\bf Abstract}
\end{center}
\noindent

In this work, we study the end-of-the-world (EOW) branes anchored to the boundaries of BCFT${}_2$ dual to the BTZ black hole. First, we explore the thermodynamics of the boundary system consisting of the conformal boundary and two EOW branes. This thermodynamics is extended by the tension appearing as the effective cosmological constant of JT black holes on the EOW branes. The tension contribution is identified with the shadow entropy equivalent to the boundary entropy of the BCFT${}_2$. The thermodynamics of the JT black holes and the bulk of BCFT${}_2$ can be combined into a novel grafted thermodynamics based on the first law. Second, we focus on the observer's view of the EOW branes by lowering the temperature. We show that the EOW branes generate a scale called ``reefs" inside the horizon. This scale also appears in the grafted thermodynamics. At high temperatures, observers on the EOW branes see their respective event horizons. The reef starts to grow relatively to the horizon size at the temperature, $T_{grow}$. As the temperature cools down the reef area fills the entire interior of the JT black holes at the temperature $T_{out}$. Then, the observers recognize their horizons disappear and see the large density of the energy flux. At this temperature, the two JT regions become causally connected. This connected spacetime has two asymptotic $AdS_2$ boundaries with a conformal matter. Also, we comment on the grafted thermodynamics to higher dimensions in Appendix B.  
\\ [15mm]

\vspace{5mm}
\end{titlepage}

%%%%%%%%%%%%%%%%%%%%%%%%%%% BODY OF PAPER %%%%%%%%%%%%%%%%%%%%%%%%%%%
\baselineskip 6.6mm
\renewcommand{\thefootnote}{\arabic{footnote}}
\setcounter{footnote}{0}

%%%%%%%%%%%%%%%%%%%
\section{Introduction}
%%%%%%%%%%%%%%%%%%%

The information problem based on Hawking radiation \cite{Hawking:1975vcx, Hawking:1976ra} is essential to understanding quantum gravity. The unitarity of the system requires a sudden change of the entropy curve at a horizon scale \cite{Page:1993df, Page:1993wv}. Recently, there has been a prominent development in this problem \cite{Almheiri:2019hni, Almheiri:2019qdq, Penington:2019kki, Almheiri:2020cfm}. Thanks to this development, we can see how the entropy curve can be obtained, and the origin of the dramatic change of the entropy is coming from a quantum wormhole in a replica space \cite{Almheiri:2019qdq} and the entanglement between radiation outside of the horizon and interior region \cite{Almheiri:2019hni}. The next step of this success would be the physical implication of the sudden change of the black hole's interior. This work is an endeavor to find the physical aspect of this change inside the horizon.

The modern approach to obtain the entropy curve, the so-called ``Page curve", is inspired by a holographic consideration of the entanglement \cite{Maldacena:2013xja, Almheiri:2019hni, Park:2022abi}. A construction of the model is made of the EOW branes and $AdS_3$ bulk for a pure system. This EOW brane or the Planck brane is also used to construct a holographic setup of the boundary conformal field theory (BCFT) \cite{Takayanagi:2011zk, Fujita:2011fp, Suzuki:2022xwv}. Therefore, one may say that the holographic BCFT is also important to understand quantum gravity. See ,{\it e.g.}, \cite{Erdmenger:2014xya, Geng:2020qvw, Geng:2020fxl, Akal:2020wfl, Akal:2020twv, Chen:2020uac, Chen:2020hmv, Hernandez:2020nem, Geng:2021mic, Geng:2022dua, Grimaldi:2022suv, Suzuki:2022xwv, Kawamoto:2023wzj}. Also, there are recent applications of BCFT or EOW brane for the early universe or swampland criteria. See, {\it e.g.}, \cite{Antonini:2019qkt, Rozali:2019day, Park:2020jio, Park:2020xho, VanRaamsdonk:2021qgv, Park:2021wep, Antonini:2022blk, Ross:2022pde, Geng:2023iqd}.

Although the boundary is ignored usually for simplicity, it is of great importance in physics. Physicists usually allude to its existence to consider realistic situations. The reason why such an important concept is ignored is that this breaks some symmetries such as translation. Due to this symmetry reduction, one may have some trouble dealing with physical problems under consideration. One way to alleviate the difficulty is imposing a suitable boundary condition that does not spoil all the bulk symmetry. Even in a $d$-dimensional conformal field theory (CFT), it is an intriguing problem to introduce boundaries. Since CFTs have a huge symmetry, there is hopefully a possibility to find boundary conditions that make field theory calculation doable or integrable. Such boundary conditions are, of course, subject to bulk conformal symmetry. The existence of boundaries breaks the conformal symmetry from $SO(2,d)$ to $SO(2,d-1)$. Cardy studied the systematic construction of boundaries in the two-dimensional CFT, and he explains how to construct consistent boundary conditions \cite{Cardy:1984bb}. See also \cite{McAvity:1995zd}. This construction defines BCFT as consisting of the bulk and boundary degrees of freedom.

An EOW brane is proposed to introduce a boundary point at the boundary system in holographic BCFT setup \cite{Takayanagi:2011zk} and the properties of this system are widely studied. However, the EOW brane itself can describe the gravity system with a CFT matter. In this work, we focus on the perspectives of the observers living on the EOW branes. The induced metric on the EOW brane provides the bulk geometry of an observer on the brane. Our construction is based on a thermal state, so the BTZ black brane or black hole will be considered as a background. In this setup, the induced metric of the EOW brane is given by a JT black hole with the same temperature as the BTZ black hole. Thus we have two kinds of thermodynamics in a total system. This thermodynamics has an additional parameter from the tension of the brane. Therefore, it is legitimate to vary the tension in the total thermodynamic system. The asymptotic boundary of the BTZ black hole and the EOW brane make a shadow on the horizon. This horizon area increased by the tension is naturally interpreted as a thermal entropy of the total BCFT system. In addition, it can be easily shown that the increased shadow entropy by the tension is the same as the boundary entropy of the dual BCFT \cite{Takayanagi:2011zk}.

Let us look at the EOW brane thermodynamics. There are two EOW branes dual to the boundaries of the BCFT${}_2$. The thermodynamics of each EOW brane is described by a JT black hole\footnote{See \cite{Geng:2022slq, Geng:2022tfc} for a similar JT black hole configuration on the Karch-Randall brane \cite{Karch:2000ct, Karch:2000gx}.} with a tension parameter changing the cosmological constant effectively. This is reminiscent of the ``thermodynamic volume" study of black holes \cite{Dolan:2011xt, Kubiznak:2012wp, Kubiznak:2016qmn}. Therefore, we may extend the thermodynamics with the thermodynamic volume and pressure. Also, this can be related to the boundary entropy leading to a new kind of temperature as a thermodynamic conjugate variable. On the other hand, the total system has already the thermodynamics of the BTZ system. However, the horizon area is increased by the tension. Comparing these two thermodynamics, we obtain a first law combining these thermodynamics. We name this novel thermodynamics ``grafted thermodynamics" based on a first law. This first law has a factor connecting the bulk thermodynamics of the BCFT and the JT systems. We call this factor a ``wrapping factor" playing an important role in the interior of the horizon.

After constructing the grafted thermodynamics, we focus on the viewpoint of observers sitting on the boundary of the JT black holes. We describe what happens under the temperature lowering of the BTZ black hole. At high temperatures, the EOW brane observers feel each horizon on each brane. In addition, the EOW branes can dive into the interior of the horizon. Each brane can be stretched toward the singularity of the BTZ black hole. However, they touch each other before meeting the singularity inside the horizon. From the perspective of the observers, this can be regarded as a region surrounding the singularity. We call this region the ``reef" because it has a peculiar behavior of lowering temperature. Below a certain temperature $T_{grow}$, this reef region rapidly grows and finally touches the horizon at a temperature $T_{out}$.

More lowering of the temperature leads to a single EOW brane without a horizon. In this stage, there is excited energy density due to an initial singular brane configuration. This can be regarded as a matter created by a geometric energy. In this era, the observers on the EOW brane see a connected spacetime with two asymptotic boundaries ($AdS_2$). There is a gas of the conformal matter in this spacetime. As a result, the two branes anchored to the boundaries of the BCFT merge into one single brane. See \cite{Geng:2021iyq, Miyaji:2022dna, Biswas:2022xfw} for recent studies about merging EOW branes.

It would be interesting to extend this toy evolution in higher dimensions. In higher dimensions, richer structures are possible. For instance, one can consider the ball shapes of the EOW brane, and the electric or magnetic charge and angular momentum can be introduced to the EOW brane. We leave these generalizations as our next research project. In this work, we just provide a comment on higher dimensional cases in Appendix B.

This paper is organized as follows. Section 2 explains the holographic setup of the BCFT${}_2$ system with two disconnected EOW branes. Also, we show how to obtain the grafted thermodynamics. Section 3 considers the temperature evolution of two EOW branes. This is a procedure from two disconnected JT black holes to a horizonless spacetime. Section 4 provides a discussion of our results and future directions. In the Appendix, we show the equivalence between the boundary entropy and the shadow entropy in two dimensions. Also, we provide a speculation for general grafted thermodynamics in higher dimensions.

%%%%%%%%%%%%%%%%%%%%%%%%%
\section{Grafted Thermodynamics of Holographic BCFT${}_2$}
%%%%%%%%%%%%%%%%%%%%%%%%%

In this section, we build a boundary conformal field theory with temperature in two dimensions using a holographic approach \cite{Takayanagi:2011zk}. Also, we construct a first law describing the grafted thermodynamics.

%%%%%%%%%%%%%%%%%%%%%%%%%
\subsection{EOW branes for BCFT${}_2$ with temperature}
%%%%%%%%%%%%%%%%%%%%%%%%%

\begin{figure}[h!]
    \centering
    \includegraphics[width=10cm]{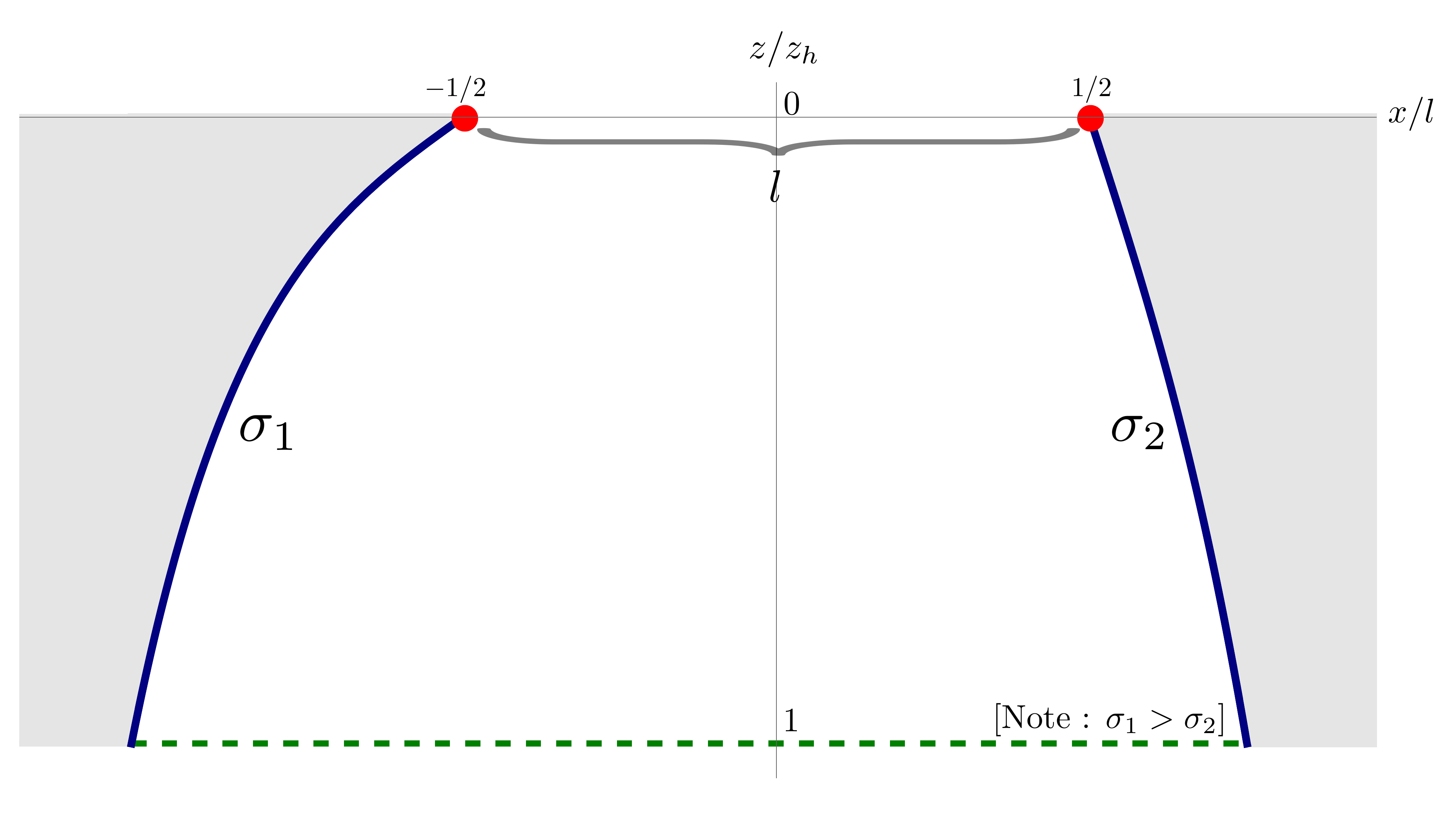}
    \caption{The indigo curves depict the EOW branes. These locations of the left and right branes are denoted by $b_1$ and $b_2$ in (\ref{t action}), respectively. The red points are the boundaries of BCFT${}_2$. $\mathcal{M}$ in the action (\ref{t action}) is the white region, and the dashed line stands for the horizon. The straight line between the red points is the spatial volume of the BCFT${}_2$, which is the spatial part of $\partial\mathcal{M}$ in (\ref{t action}). Its size is $l=\mathcal{V}_1$.}
    \label{fig:Cartoon01}
\end{figure}

We follow the proposal of \cite{Takayanagi:2011zk} starting with
\begin{align}\label{t action}
S =& \frac{1}{16\pi G_{(3)}} \int_{\mathcal{M}} d^3x\sqrt{-g}\left(\mathcal{R}-2\Lambda\right) + \frac{1}{8\pi G_{(3)}} \sum_{i=1}^2 \int_{{b}_i} d^2x \sqrt{-h} \left(K + \mathcal{L}_{b_i}\right)\nonumber\\
&+\frac{1}{8\pi G_{(3)}} \int_{\partial\mathcal{M}} d^2 x\sqrt{-\gamma}\left(K +\mathcal{L}_{c.t}\right)\,,
\end{align}
where $\mathcal{M}$ is the bulk surrounded by the conformal boundary denoted by $\partial\mathcal{M}$, the horizon and two EOW branes. See Figure \ref{fig:Cartoon01}. $b_i$ denotes the locations of the End of the World (EOW) branes. Also, $\mathcal{L}_{b_i}$ is the Lagrangian density of the matter on the EOW branes. The $K$ is the Gibbons-Hawking term on each boundary. $\mathcal{L}_{c.t}$ denotes the counter Lagrangian for the holographic renormalization. The bulk geometry is determined by the Einstein equation:
\begin{align}\label{eom00}
\mathcal{R}_{MN}-\frac{1}{2}g_{MN}(\mathcal{R} - 2 \Lambda ) = 0\,,
\end{align}
where $\Lambda=-1/L^2$. We assume that the back reaction of the EOW branes is negligible. Otherwise, the right-hand side does not vanish. We maintain this probe approximation throughout this paper.

As a background geometry, we take the BTZ black brane given by the following metric:
\begin{align}\label{BackMetric}
ds^2 = - \frac{f(z)}{z^2} dt^2 + \frac{dx^2}{z^2} + L^2 \frac{dz^2}{z^2 f(z)}\,,
\end{align}
where $z$-coordinate is dimensionless and $f(z)$ is 
\begin{align}\label{f00}
f(z) =  1 - m L^2 z^2~.
\end{align}
The $m$ denotes the black hole mass parameter related to the location of horizon $z_h$ as $m=1/(z_h^2 L^2)$. The asymptotic boundary of this black brane is sitting on $z=0$. Since we consider this planar black brane(or black string), the range of $x$-coordinate is infinite. Thus the dual field theory is defined on $\mathbb{S}^1\times\mathbb{R}$, where $\mathbb{S}^1$ is the temperature circle. On the other hand, the BTZ black hole case describes the field theory on a space $\mathbb{S}^1$. In this case, $x=R\,\theta$ ranges from $0$ to $2\pi R$. We will come back to this case later.

In the aforementioned probe approximation, the locations of the EOW branes $b_i$'s are determined by the background geometry. To determine the locations, we need to solve a junction condition in terms of a unit normal vector to each brane. If we denote the normal vector by $n_M$, then the induced metric and the extrinsic curvature can be written as
\begin{align}
h_{MN}= g_{MN}-n_M n_N~,~K_{MN} =  \frac{1}{2}\left(\nabla_M n_N+\nabla_N n_M\right)\,.
\end{align}
In addition, the scalar extrinsic curvature is
\begin{align}
K=h^{MN}K_{MN}=\nabla^M n_M\text{, where}~~h^{MN} = g^{MN}-n^M n^N~.
\end{align}

The location of each EOW brane is determined by a function $g_i(z)$. For the brane on the right in Figure \ref{fig:Cartoon01}, the location is given as $x=L g_2(z)$. Therefore, we introduce a coordinate $y=L\, g_2(z)-x$ for the brane together with $dy=L\, g'_2(z) dz-dx$. One can easily apply the same procedure to the opposite brane, so we drop the subscript $i$. Since we take the inward direction for each normal vector, the $y$-direction is nothing but the direction of $n_M$. Then, the background metric (\ref{BackMetric}) can be decomposed by
\begin{align}
ds^2 = N^2 dy^2 + h_{ab}\left(d\sigma^a + \mathcal{N}^a dy \right)\left(d\sigma^b + \mathcal{N}^b dy \right)\,. 
\end{align}
Here, the brane-surface coordinates are $\sigma^a=(t,z)$. The induced metric, the shift vector, and the lapse function are given as follows:
\begin{align}\label{Induced-Metric}
h_{ab} = \text{diag}\left( -\frac{f}{z^2},\, L^2 \frac{1+ f g'^2}{z^2 f} \right) ~~,~\mathcal{N}^a=(0,\, -\frac{f g'}{L\left(1+ f g'^2\right)}  )~,~N=\frac{1}{z\sqrt{1 + f g'^2}}\,.
\end{align}
Using this decomposition, the variation of the action is given by $\mathcal{X}^{ab}\delta h_{ab}$ which should vanish. Here $\mathcal{X}^{ab}$ is some expression in terms of $h_{ab}$, $N$, $\mathcal{N}_a$ and $\mathcal{L}_{b_i}$. As we mentioned, the locations of the EOW branes are not fixed. Therefore, the induced metric $h_{ab}$ can vary. This requires $\mathcal{X}_{ab}=0$ rather than vanishing $\delta h_{ab}$. This is the junction condition given as follows:
\begin{align}\label{Junction00}
\mathcal{X}_{ab}=- K_{ab} + \left(  K - \mathcal{L}_{\mathcal{B}_i} \right)h_{ab} - 2 \,\frac{\delta\mathcal{L}_{b_i}}{\delta h^{ab}}=0\,,
\end{align}
where the extrinsic curvature can be written in terms of the metric decomposition as
\begin{align}
K_{ab} = \frac{1}{2N}\left(\partial_y h_{ab}-\nabla_a^{[h]}\mathcal{N}_b-\nabla_b^{[h]}\mathcal{N}_a \right)~,~~K = h^{ab}K_{ab}\,.
\end{align}
Therefore, one can find the brane location by solving (\ref{Junction00}).

Now, we may choose the matter localized on the EOW branes. One of the simplest matters is the constant tension without other energy excitation on the branes. This constant tension satisfies the null energy condition. Thus we choose $\mathcal{L}_{b_i}=-\sigma_i$, where we denote again the EOW brane index by $i$. Then the only relevant equation of (\ref{Junction00}) is\footnote{To show this, one may use the Einstein equation.} 
\begin{align}\label{JC_dg}
g_i'(z)= \frac{L \sigma_i}{\sqrt{1-f(z)L^2 \sigma_i^2}}\,.
\end{align}
Plugging (\ref{f00}) into this equation, one can find the embedding function as follows:
\begin{align}
g_i(z) = \frac{1}{L \sqrt{m}}\tanh ^{-1}\left(\frac{L^2 \sqrt{m} \sigma_i  z}{\sqrt{1-L^2 \sigma_i ^2 f(z)}}\right)+g_i(0)\, ,
%\frac{1}{L\sqrt{m}} \tanh ^{-1}\left(\frac{ \sqrt{m}\,L^2 \sigma_i z}{\sqrt{1-L^2 \sigma_i^2+ m L^4 \sigma_i^2 z^2}}\right) + g_i(0)\,,
\end{align}
where the additive constant $g_i(0)$ respects the translation symmetry along the $x$-direction of the background geometry. This configuration describes $i$-th brane with tension $\sigma_i$ touching one of the BCFT boundaries and the outside of the branes is not included as the gravity dual to the BCFT. We plot a configuration\footnote{One can find the same configuration in \cite{Geng:2021iyq}.} in Figure \ref{fig:Cartoon01}.

%%%%%%%%%%%%%%%%%%%%%%%%%
\subsection{Thermodynamics of BCFT${}_2$ with tension}
%%%%%%%%%%%%%%%%%%%%%%%%% 

In this subsection, we discuss the thermodynamics of the BCFT${}_2$. Before considering the full system, we first take into account the thermodynamics of the BTZ black brane. This BTZ black brane has thermodynamics supported by a black hole parameter variation. This is nothing but the first law of the black brane with the size $\mathcal{V}_1$. The expression is
\begin{align}\label{1st-Law}
\delta\left( \frac{m L}{16\pi G} \mathcal{V}_1 \right)  = T \delta\left(\frac{\mathcal{A}_H}{4G} \right) - \frac{m L}{16\pi G} \delta\mathcal{V}_1
\end{align}
where $m$, $T$, $\mathcal{A}_H$ and $\mathcal{V}_1$ are the black hole mass parameter, the Hawking temperature, the horizon area, and the size of the system, respectively. To show the above relation, we used the following expressions:
\begin{align}
\mathcal{A}_H=\frac{\mathcal{V}_1}{z_h}~,~T = \frac{1}{2\pi L z_h} \,.
\end{align}
The black brane mass parameter is given by the horizon condition $f(z_h)=0$, which turns out to be the Smarr relation.

Using the total mass $M\equiv \frac{m L}{16\pi G} \mathcal{V}_1$ and the Bekenstein-Hawking entropy, (\ref{1st-Law}) can be written as follows:
\begin{align}
dM = T \delta \mathcal{S}\,,
\end{align} 
where the entropy is defined as $\mathcal{S}\equiv\frac{\mathcal{A}_H}{4G} $. Also, the Smarr relation is given by 
\begin{align}
2 M= T \mathcal{S}\,.
\end{align}
Applying the AdS/CFT correspondence, one can interpret this system as a strong coupling limit of a two-dimensional CFT on $\mathbb{R}$. The horizon area is the same as the thermal entropy of this CFT. This infinite system has entropy density and energy density in terms of the black brane parameters:
\begin{align}
\epsilon=\frac{m L}{16\pi G_{(3)}}~,~s=\frac{\mathcal{A}_H}{4G_{(3)}\mathcal{V}_1}~\,.
\end{align}
Then, the Smarr relation and the first law can be written in terms of the dual field theory variables as
\begin{align}\label{1st-Law-FT}
\mathcal{F}= -\mathcal{P}=\epsilon - s T~,~\delta E_B= T \delta \mathcal{S}-\mathcal{P}\delta \mathcal{V}_1\,,
\end{align}
where the energy density $\epsilon$ and the pressure $\mathcal{P}$ are given by components of the holographic energy-momentum tensor. Also, the free energy density $\mathcal{F}$ can be obtained by the Euclidean on-shell action. $E_B$ denotes the energy for the space $\mathcal{V}_1$, i.e., $E_B=\epsilon \mathcal{V}_1$. The first equality is guaranteed by the translation symmetry along $x$-direction. Also, one can show that $\epsilon=\mathcal{P}$. This is consistent with the conformal symmetry.

Now, let us look at Figure \ref{fig:Cartoon01}. The bulk dual to the BCFT${}_2$ system is depicted by the white region. This region is surrounded by the conformal boundary, two EOW branes and the horizon. Therefore, it is natural to take the total thermal entropy as the horizon area touched by the EOW branes. This can be written as
\begin{align}
\mathcal{S}_{\text{total}} &= \mathcal{S} + \mathcal{S}_b = \frac{1}{4G_{(3)}}\left(\frac{\mathcal{V}_1}{z_h} +\sum_{i=1}^2  \frac{L}{z_h} g_i(z_h)\right)\nonumber\\
&=\frac{1}{4G_{(3)}}\left(\frac{\mathcal{V}_1}{z_h} + \sum_{i=1}^2 L \tanh^{-1} L\sigma_i \right)\,,
\end{align}
where the allowed range of the tension $\sigma_i$ is $0< \sigma_i < 1/L$. We call $\mathcal{S}_b$ the shadow entropy, and it is known that this shadow entropy is the same as the boundary entropy of the BCFT${}_2$ \cite{Takayanagi:2011zk}. The $i$-th boundary entropy $\mathcal{S}_b^{(i)}$ is related to the g-function $\mathbf{g}_i$ as $\mathcal{S}_b^{(i)} = \log\mathbf{g}_i$. \footnote{ Here, it is notable that there is no guarantee for the equivalence between the holographic boundary entropy and the shadow entropy in the higher dimensions or black holes with another conserved charges, such as, electric charge and angular momentum. We checked that they are not identical even in the one-higher dimensions with a simple extension of the EOW brane.}

Also, the temperature times the variation of entropy takes the following form\footnote{  Using $\tau = \frac{\sigma_i L}{\sqrt{1-\sigma_i^2 L^2}}$, the last term can be written as $\sum_{i=1}^2\frac{L \delta \sigma_i }{8 \pi  G_{(3)} z_h \left(1-L^2 \sigma_i^2\right)
}=\sum_i \frac{\delta \tau_i}{8\pi G_{(3)}L z_h \sqrt{1 +\tau_i^2}}$. }:
\begin{align}\label{TdS_tot}
T \delta \mathcal{S}_{\text{total}} = -\frac{\mathcal{V}_1  \delta z_h}{8 \pi  G_{(3)} L z_h^3}+ \frac{\delta \mathcal{V}_1}{8 \pi  G_{(3)} L z_h^2}+\sum_{i=1}^2\frac{L \delta \sigma_i }{8 \pi  G_{(3)} z_h \left(1-L^2 \sigma_i^2\right)}\,.
\end{align}
Using the expression of energy and pressure, one can rewrite the above variation as
\begin{align}\label{first law 00}
\delta E_B + \sum_{i=1}^2 \frac{1}{8\pi G_{(3)} z_h}\delta  (\tanh^{-1} L \sigma_i) =T\delta \mathcal{S}_{\text{total}}-\mathcal{P}\delta \mathcal{V}_1\,.
\end{align}
This is an extended first law of the BCFT${}_2$ with the tension $\sigma_i$'s, which must be associated with the boundary conditions or the boundary states of BCFT${}_2$. The boundary points are also the boundaries of the EOW branes. So one can apply the second holography to these boundary points. Then, the boundary system can be regarded as a system composed of CFT${}_2$ with two quantum mechanics at the endpoints of the CFT volume.

\subsection{Grafted thermodynamics with EOW branes}

Let us start with the induced metric on an EOW brane (\ref{Induced-Metric}):
\begin{align}\label{metric-g}
ds^2 &= - \frac{f(z)}{z^2}dt^2 + \frac{L^2}{z^2} \frac{1+f(z) g_i'(z)^2 }{f(z)} dz^2\nonumber\\
&= - \frac{f(z)}{z^2}dt^2 + \frac{1}{z^2 \sigma_i^2} \frac{dg_i^2 }{f(z)}\,,
\end{align}
where we used the junction condition (\ref{JC_dg}) for the last equality. This metric expression can be written as follows:
\begin{align}
ds^2 = \frac{1}{X^2} \left[ - \frac{\tau_i^2}{z_h^2}\left(1- \frac{X^2}{\tau_i^2}\right)dt^2 + L^2 \frac{1+ \tau^2}{1+X^2}\frac{dX^2}{1-\frac{X^2}{\tau_i^2}} \right]\,,
\end{align}
where
\begin{align}
\tau_i=\frac{\sigma_i L}{\sqrt{1-\sigma_i^2 L^2}}~~,~~X = \sinh\left(\frac{g_i}{z_h}\right)\,.
\end{align}
We will drop the subscript to concentrate on one single EOW brane for the time being.

This two-dimensional geometry has the horizon at $X=\tau$. The temperature is the same as the temperature of the BTZ black hole. Of course, the temperature is independent of the tension $\sigma$. Also, the scalar curvature of this geometry is given by $\mathcal{R}^{(2)}= - \frac{2}{L^2 (1+ \tau^2)}$. 
 Thus, this black hole geometry can be mapped into the black hole in the 2-dimensional JT gravity. To realize this mapping, one may take a coordinate transformation that leads to the following metric form: 
\begin{align}\label{metic2D}
ds^2 = - A(y) dt^2 +  L^2(1+ \tau^2)  \frac{dy^2}{B(y)}\,
\end{align} 
with
\begin{align}\label{induced BH}
A(y) = B(y) = y^2 - \left(\frac{1+\tau^2}{z_h^2} \right)~,~y=\frac{\tau}{{z_h}}\frac{\sqrt{1+X^2}}{X}\,.
\end{align}
This metric can be embedded as a solution of the JT gravity whose action is given by
\begin{align}\label{action JT}
\mathcal{I}_{\text{JT}} = \frac{1}{2} \int d^2x \sqrt{-h} \,\phi \left( R + \frac{2}{L^2(1 + \tau^2)} \right)\,.
\end{align}
Plugging (\ref{metic2D}) into this action, we can obtain the following reduced action: 
\begin{align}
\mathcal{I}_{\text{JT}} =   \int d^2 x \frac{\phi  \left(4 A^2+B \left(A'\right)^2-A \left(2 B A''+A' B'\right)\right)}{4 A^{3/2} \sqrt{B} L \sqrt{\tau ^2+1}}\,.  
\end{align}
We may take $A$ as $B$ by a suitable coordinate transformation. The equations of motion for $A$, $B$ and $\phi$ with the gauge $B=A$ are given by
\begin{align}
2 \phi - \phi' A' =0~~,~~A''= 2\,.
\end{align}
Also, one may take a further gage choice, $\phi= y$. Then, we have 
\begin{align}
A(y) = y^2 - y_h^2\,
\end{align}
where the $y_h$ is the location of the horizon. Suppose there is an observer confined on this EOW brane and such an observer can't recognize the bulk geometry. Then, the induced metric is the only observable to this observer. Therefore, the observer regards this EOW brane as a world of the JT gravity effectively. This is the reason why we introduce the map from the induced metric to the JT black hole. It is notable that this mapping does not tell us equivalence between the boundary action in (\ref{t action}) and the action of the JT system.

Now, let us consider the thermodynamics of this JT black hole. The temperature of this black hole and the entropy can be derived using the induced metric. The obtained result is
\begin{align}
T = \frac{A'(y_h)}{4\pi L\sqrt{1+\tau^2}}= \frac{y_h}{2\pi L\sqrt{1+\tau^2}}~~,~~S_{\text{JT}}=2\pi \phi(y_h)= 2\pi y_h \,.
\end{align}
One can check that this temperature is nothing but the BTZ black brane temperature ($T={1}/{2\pi z_h L}$) using (\ref{induced BH}). Also, the integration constant $y_h^2$ is identified with the holographic energy of the JT gravity as follows:
\begin{align}
E_{\text{JT}}= \frac{y_h^2}{2 L\sqrt{1+\tau^2}}\,.
\end{align}  
One can notice that the first law of thermodynamics, $dE_{JT}= T dS_{\text{JT}}$, is satisfied when $\tau$ is constant. If we allow the variation of $\tau$, the first law becomes a different form. Since $L^2(1+\tau^2)$ provides an effective cosmological constant, the variation of the tension is equivalent to the variation of the cosmological constant for the JT black hole. This consideration is reminiscent of studies of the thermodynamic volume \cite{Dolan:2011xt, Kubiznak:2012wp, Kubiznak:2016qmn}. The variation of $\tau$ leads to
\begin{align}
\delta E_{\text{JT}} = T \delta S_{JT}-\frac{y_h^2 \tau \delta \tau}{2L(1+\tau^2)^{3/2}}\,.
\end{align}
The last term is the effect of the varying cosmological constant. In higher dimensions, the cosmological constant can be regarded as a kind of pressure $\mathcal{P}_\tau=  1/L^2(1+\tau^2)$. Then, the last term becomes $\mathcal{V}_\tau \delta \mathcal{P}_\tau$. The explicit expression is
\begin{align}
\mathcal{V}_\tau \delta \mathcal{P}_\tau = \left(\frac{y_h^2 L \sqrt{1+\tau^2}}{4} \right) \delta  \left(\frac{1}{L^2(1+\tau^2)} \right)\,.
\end{align}
The conjugate volume $\mathcal{V}_\tau$ is called the thermodynamic volume in literature. See \cite{Kubiznak:2012wp, Gunasekaran:2012dq, Kubiznak:2016qmn, Altamirano:2014tva} for instance. Interestingly, this is related to the volume inside the horizon as follows:
\begin{align}
\mathcal{V}_\tau = \frac{1}{2}\int_0^{y_h} dy \sqrt{-h}\, \phi\,.
\end{align}
More correctly, this quantity is equivalent to the average position of the EOW brane inside the horizon with some numerical factor because the dilaton is identified with the coordinate $y$. Using these pressure and thermodynamic volume, we can write down the thermodynamic first law as follows:
\begin{align}\label{1st law JT}
\delta E_{JT} = T \delta S_{JT} + \mathcal{V}_\tau \delta  \mathcal{P}_\tau \,.
\end{align}
Therefore, the mass or holographic energy of the JT black hole is not the internal energy but the enthalpy. The internal energy is $U_{JT}=E_{JT}- \mathcal{P}_\tau \mathcal{V}_\tau=E_{JT}/2$ satisfying $\delta U_{JT}= T \delta S_{JT}-\mathcal{P}_\tau \delta \mathcal{V}_\tau$.

These thermodynamic volume and pressure are introduced by varying the tension parameter $\tau$. This varying tension also induces a variation of the boundary entropy. If, thus, we express the last term of the first law (\ref{1st law JT}) in terms of the boundary entropy and its thermodynamically conjugate parameter, then the first law becomes 
\begin{align}
\delta E_{JT} = T \delta S_{JT} - \hat{T} \delta \mathcal{S}_b\,,
\end{align}
where $\mathcal{S}_b$ is the boundary entropy and shadow entropy in this case. The thermodynamic temperature $\hat{T}$ conjugate to $\mathcal{S}_b$ is given by
\begin{align}
\hat{T}=  \frac{4\pi\tau G_{(3)}}{z_h L}   \, T\,.
\end{align}
This quantity depends on temperature and tension.

If we find the first law of the total system (\ref{first law 00}) in terms of the JT black hole parameters, we arrive at the following relation:
\begin{align}\label{first law 01}
\delta  E_B - T \delta \mathcal{S}_{total} +\mathcal{P} \delta \mathcal{V}_1 = \sum_i^2 \mathbf{w}_i  \left(  \delta E^i_{JT}- T \delta S^i_{JT} \right)  \,,
\end{align}
where $\mathbf{w}_i\equiv\frac{T}{\hat{T}_i}$. One can see that the thermodynamics of two JT black branes can be incorporated into the thermodynamics of CFT${}_2$ with volume $\mathcal{V}_1$ by a ratio between two temperatures. We call this factor $\mathbf{w}_i$ ``wrapping factor". This factor controls the strength of the interplay between two thermodynamics.

\section{EOW Perspective to Thermal $BCFT_2$ on $\mathbb{S}^1$}

In this section, we consider the BTZ black hole with a finite $x$-range from $0$ to $2\pi R$. Also, we trace out the observer's view on each EOW brane while lowering the temperature.

\subsection{Matter excitation by fusion of EOW branes}

Since the BCFT${}_2$ has two boundaries, it is hard to see whether the original CFT without boundaries is defined on $\mathbb{S}^1$ or $\mathbb{R}$. However, the gravity duals of both cases are different. In addition, black brane solutions are usually taken as limiting solutions of black hole geometries. Therefore, it may be natural to regard the BTZ black hole as a gravity dual. So, this section will be devoted to seeing what happens when we consider the BTZ black hole as the background geometry.

 \begin{figure}[h!]
    \centering
    \includegraphics[width=8cm]{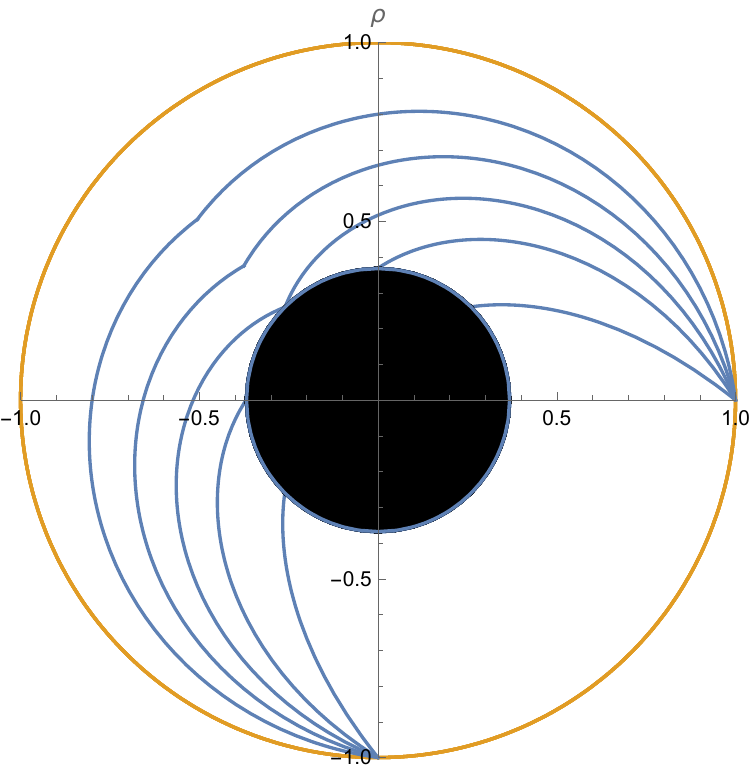}
    \caption{EOW brane locations under lowering the temperature: We choose the bulk of BCFT${}_2$ regions to be placed in $-\frac{\pi}{2}< \theta< 0$, so $\mathcal{V}_1=l = \frac{\pi R}{2}$. Two EOW branes have the same tension with $\sigma_1=\sigma_2=\frac{1}{3 L}$ for simplicity. The temperature is taken as $T R = \left\{0.07, 0.035, T_{out}={2 \tanh ^{-1}\left(\frac{1}{3}\right)}/{(3 \pi ^2)}, 0.015, 0.008 \right\}$. As the temperature lowers, the shadow entropy becomes larger. The EOW branes touch the horizon directly in high temperatures. At $T=T_{out}$, two EOW branes touch each other at a horizon point. Below this temperature, two EOW branes touch outside the horizon.}
    \label{fig:EOWs01}
\end{figure}

We may use the same junction equation (\ref{JC_dg}) in this case. As we discussed earlier, the only thing we have to take into account is to choose an angle-coordinate $\theta$ with $x=R \theta$. Thus the BCFT system size $\mathcal{V}_1$ is smaller than $2\pi R$. To visualize the EOW configuration, we take a coordinate $\rho=e^{-z/z_h}$. In addition, we can relate the location of an EOW brane as $\theta_i= \frac{L}{R}\left(g_i(z)-g_i(0)\right)$. 
Then, the i-th EOW brane location in this coordinate is determined by
\begin{align}
\rho=\exp\left\{-\frac{z}{z_h}\right\} = \exp\left\{-\frac{\sqrt{1- \sigma_i^2 L^2}}{\sigma_i L} \sinh (2\pi RT \theta )\right\}\,.
\end{align}
We plot the EOW brane locations for a certain tension $\sigma_i$ by lowering the temperature in Figure \ref{fig:EOWs01}. For simplicity, we choose the same tension for both EOW branes from now on. As one can see in the figure \ref{fig:EOWs01}, the EOW branes come out from the horizon area below a temperature denoted by $T_{out}$. This temperature is determined by 
\begin{align}\label{Temp-out}
T_{out} = \frac{1}{2\pi  (2\pi R- \mathcal{V}_1)} \sum_{i=1}^2   \tanh^{-1} L \sigma_i\,.
\end{align}
Below this peculiar temperature, two disconnected EOW branes are combined into one single brane with a cusp. Due to this cusp, these configurations are not physical solutions. A more physically relevant situation is a smooth EOW brane with some matter distribution. Let us look at the initial cusp configuration at $T=T_{out}$. This singular configuration causes an infinite energy density at the merging point. Therefore, this large extrinsic curvature should be compensated by an infinite energy density at the cusp point of the single brane. By this reasoning, we will introduce the matter on this single brane, and show the initial cusp configuration can be replaced with an EOW brane accompanied by the infinite energy density. Such a configuration including matter distribution will be found in the next subsection and replace these cusp-solutions without matter.

The excited matter on the EOW brane is described by a nonvanishing energy-momentum tensor, which changes the junction  equation (\ref{Junction00}) as follows:
\begin{align}\label{junctionEq}
K_{ab} - K h_{ab} = \sigma h_{ab} + T_{ab}\,.
\end{align} 
A correct way to depict this problem after the brane collision and fusion is that find the time-dependent EOW brane configuration with an initial energy configuration located at the cusp. However, this is a difficult dynamical problem due to the time dependence of the brane location. Thus we take an easier route. We find static matter and brane configurations in temperatures, and we compare such configurations to deduce how the EOW brane evolves.

The energy-momentum tensor could be a more general form, but we chose a simple type to grab the configuration of the EOW brane. We consider a static perfect fluid as the matter on the single EOW brane. The explicit energy-momentum tensor is
\begin{align}\label{emTensor0}
T_{ab} = \epsilon(z) u_a u_b + p(z) p_{ab}\,,
\end{align}
where the $p_{ab}$ is the projection to the brane spatial direction. Also, we impose the energy-momentum conservation $\nabla_a^{[h]} T^{ab}=0$ on the EOW brane. This condition leads to
\begin{align}
p'(z)= \frac{p(z) +\epsilon(z)}{z f(z)}\,.
\end{align}
Now, the junction condition becomes
\begin{align}\label{junction eq p}
g'(z) =\frac{(\sigma +p(z))L}{\sqrt{1-f(z)(\sigma + p(z))^2 L^2}}\,.
\end{align}
Here one can notice that the denominator could vanish outside of the horizon. Such a smooth brane tip $z=z_c$ requires the following condition: 
\begin{align}
p(z_c) = \frac{1}{L}\left( \frac{1}{\sqrt{f(z_c)}} -\sigma L \right)\,.
\end{align}
The pressure on the EOW brane can be solved as follows:
\begin{align}
p(z) &= \frac{z}{\sqrt{z_h^2-z^2}} \left(  \sqrt{\frac{z_h^2}{z_c^2}-1} \, p_c +\int _{z_c}^z dw \frac{z_h}{{w}^2} \frac{\epsilon (w)}{\sqrt{1-\frac{w^2}{z_h^2}}} \right)\,,
\end{align}
where $p_c=p(z_c)$. Therefore, the pressure can be found by a given energy configuration $\epsilon(z)$. Then the modified junction condition (\ref{junction eq p}) can also be integrated.

\subsection{Conformal matter on single EOW brane}

Now let us consider a more specific case. We propose a conformal matter described by $\epsilon=p$. Then, the EOW brane configuration is determined by the following three equations:
\begin{align}\label{junction-cmatter}
\tilde{\epsilon}'(\tilde{z}) =\frac{2\tilde{\epsilon}}{\tilde{z}f(\tilde{z})}~,~ g'(\tilde{z})=\frac{z_h(\tilde{\sigma}+\tilde{\epsilon}   )}{\sqrt{1-f(\tilde{z})(\tilde{\sigma}+\tilde{\epsilon}   )^2}}~,~f'(\tilde{z})=-\frac{2(1-f(\tilde{z}))}{\tilde{z}}\,,
\end{align}
where we use dimensionless coordinate and quantities defined as $\tilde{z}=z/z_h$, $\tilde{\sigma}=\sigma L$ and $\tilde{\epsilon}=\epsilon L$ to simplify the expressions. The first equation leads to $ \tilde{\epsilon}(\tilde{z})=\frac{\epsilon_0\tilde{z}^2}{1-\tilde{z}^2}$. Plugging this energy density into the second equation, one can obtain a differential equation giving the location of the EOW brane. The third equation is nothing but the bulk Einstein equation.

The required boundary condition for the regular tip ($\tilde{z}=\tilde{z}_c$) of the EOW brane is $g'(\tilde{z}_c)=\infty$. To deal with this differential equation numerically, we find $\tilde{z}(g)$ rather than $g(\tilde{z})$. To this purpose, we consider another equivalent form of the differential equations. They are given by
\begin{align}
\tilde{\epsilon}'(\tilde{z}) = \frac{2 \tilde{\epsilon}}{Z(1-Z^2)}~,~g'(\tilde{z})=\frac{z_h (\tilde{\sigma}+\tilde{\epsilon} )}{\sqrt{1- (1-Z^2)(\tilde{\sigma}+\tilde{\epsilon}  )^2}}~,~Z'(\tilde{z})=1\,,
\end{align}
where $Z$ is nothing but $\tilde{z}$. Using the above equations, one can construct differential equations to obtain the brane location $Z(g)$ and the energy density $\tilde{\epsilon}(g)$. The set of equations is
\begin{align}\label{eom-g}
\frac{d\tilde{\epsilon}}{dg} = \frac{2 \tilde{\epsilon} }{Z(1-Z^2)} \frac{\sqrt{1- (1-Z^2)(\tilde{\sigma}+\tilde{\epsilon}  )^2}}{z_h (\tilde{\sigma}+\tilde{\epsilon} )}~,~\frac{dZ}{dg}=\frac{\sqrt{1- (1-Z^2)(\tilde{\sigma}+\tilde{\epsilon}  )^2}}{z_h (\tilde{\sigma}+\tilde{\epsilon}  )}\,.
\end{align}
The range of the parameter $g$ is from 0 to $g_c$. $g_c$ is the tip location of the EOW brane. In addition, $g=0$ is determined by $Z(0)=0$ and $\tilde{\epsilon}=0$, where the EOW brane touches the conformal boundary of the BTZ black hole. $Z(g)$ and $\tilde{\epsilon}(g)$ in this range can describe half of the brane configuration and the other half is given by the mirror copy about $g=g_c$.

 \begin{figure}[h!]
    \centering
    \includegraphics[width=10cm]{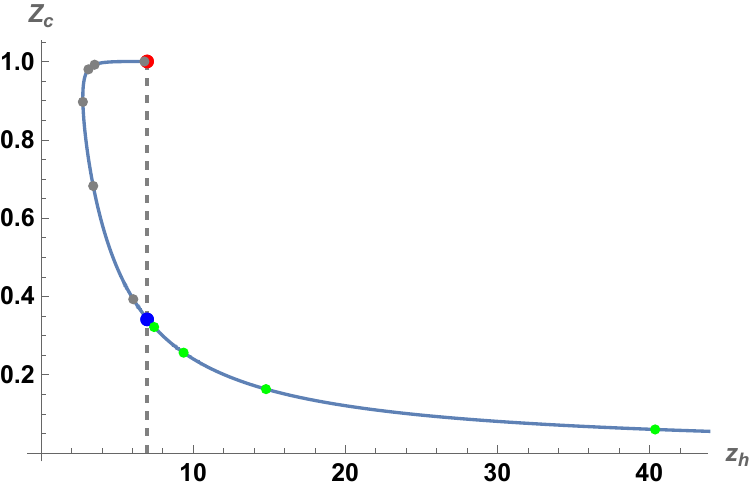}
    \caption{The possible parameter-curve $z_h(Z_c)$ for $\tilde{\sigma}=1/3$, $l=\pi R/2$}
    \label{fig:Zczh}
\end{figure}

The suitable boundary condition on the tip is required as $\frac{d\tilde{\epsilon}}{dg}|_{g=g_c} = \frac{dZ}{dg}|_{g=g_c}=0$. Due to this regularity condition, the energy density at the tip is related to the tip position as follows:  
\begin{align}
\tilde{\epsilon}(g_c)=  \frac{1}{\sqrt{1-Z_c^2}}-\tilde{\sigma} \,,
\end{align}
where $Z_c=Z(g_c)$. Therefore, the set of parameters for a solution is $\{R,~\tilde{\sigma},~z_h,~Z_c,~  g_c\}$. For given parameters, $\{R,~\tilde{\sigma},~g_c=(2\pi R- \mathcal{V}_1)/2L\}$, one can find a solution-curve $z_h(Z_c)$. We show the curve for the $\mathcal{V}_1=\pi R/2$ case in Figure \ref{fig:Zczh}. As one can see in the figure, the temperature, $T=1/(2\pi z_h L)$, is not monotonically decreasing for the fixed volume $\mathcal{V}_1$ of the BCFT${}_2$. This is quite interesting behavior. As we commented earlier, we didn't consider the time-dependent location of the EOW brane and chose the perfect fluid matter. Thus, this behavior may be regarded as an artifact of the static configuration. This is reminiscent of a first-order phase transition of D${}_7$ brane in D${}_3$ background \cite{Nakamura:2006xk}. See also \cite{Nakamura:2007nx, Seo:2008qc, Seo:2009kg, Gwak:2011wr, Gwak:2012ht, Evans:2012cx} for the related works. In our case, this is also similar to a first-order phase transition. We may expect that the EOW brane moves shortly from the red dot to the blue dot in Figure \ref{fig:Zczh} along the dashed line at the temperature $T_{out}$ (\ref{Temp-out}).

 \begin{figure}[h!]
    \centering
    \includegraphics[width=8cm]{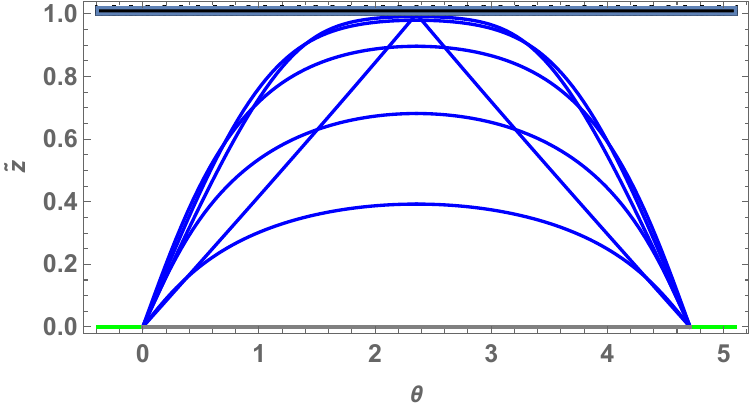}
    \includegraphics[width=8cm]{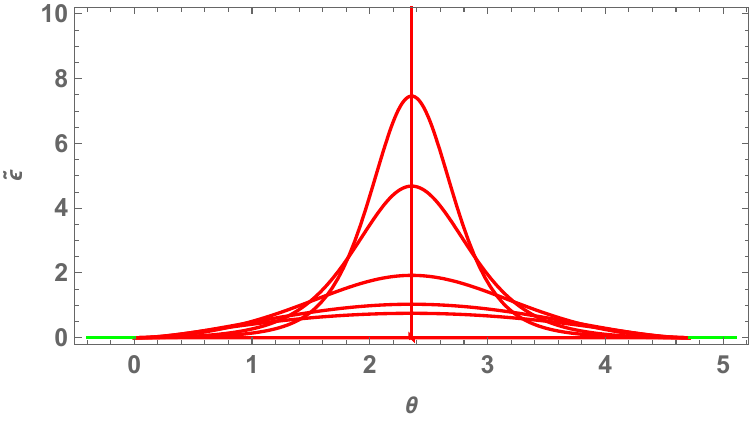}
    \caption{Initial and unphysical configurations of the EOW brane (Left) and the energy density (Right): These configurations correspond to the gray points in Figure \ref{fig:Zczh}. The black and green regions denote the horizon and the space of the BCFT. Although these configurations are unphysical, one can speculate that the extremely localized energy density can spread out toward the boundary of AdS space at the temperature $T_{out}$. The initial configuration accompanies the delta function-like energy density. This depicts red dot in Figure \ref{fig:Zczh} and replaces the brane configuration with a cusp(and without matter) at $T=T_{out}$ in Figure \ref{fig:EOWs01}.}
    \label{fig:Unphy}
\end{figure}

 \begin{figure}[h!]
    \centering
    \includegraphics[width=8cm]{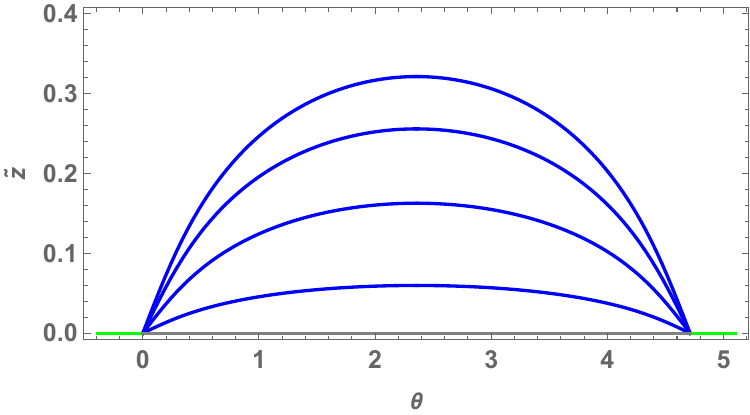}
    \includegraphics[width=8cm]{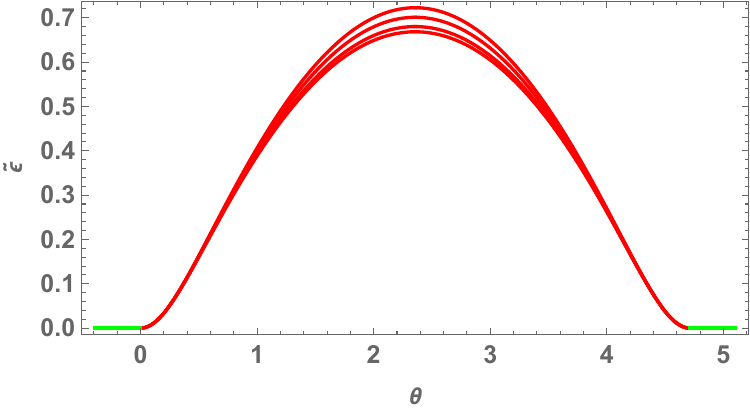}
    \caption{Physical configurations of the EOW brane (Left) and the energy density (Right): These configurations correspond to the green points in Figure \ref{fig:Zczh}. The green lines in these figures denote the space of the BCFT. One can notice that the energy density spreads out toward the boundary of AdS space by lowering the temperature.}
    \label{fig:Phy}
\end{figure}

Even though the parameter curve for solutions between the red and blue dots in Figure \ref{fig:Zczh} does not seem to be physical, it is desirable to find the solutions for this parameter curve to understand this transition. So we show the energy densities and brane locations in Figure \ref{fig:Unphy}. The provided solutions are about the gray dots in Figure \ref{fig:Zczh}. The solution for the red dot shows a triangular brane shape, almost the same as the EOW brane with a cusp. See the left figure in Figure \ref{fig:Unphy}. The energy density for this brane configuration shows a delta function-like energy density which is shown in the right figure. The other solutions of Figure \ref{fig:Unphy} show that the EOW brane becomes flattening and close to the conformal boundary of the BTZ black hole along the parameter curve from the red dot to the blue dot in Figure \ref{fig:Zczh}. Also, the right figure of Figure \ref{fig:Unphy} shows the energy density spreads out toward the conformal boundary.

The physical solutions (The green dots in Figure \ref{fig:Zczh}) are also shown in Figure \ref{fig:Phy}. The left figure shows the EOW brane gets close to the conformal boundary as the temperature decreases. The figure on the right shows the evolution of energy density. The maximum energy density decreases as the temperature lowers.

Let us focus on the observer-viewpoint in this single EOW brane. Since the Einstein tensor vanishes in two dimensions, the observers feel a geometry in terms of the scalar curvature. The induced metric is the same form as the first line in (\ref{metric-g}). Now the junction equation is the second one in (\ref{junction-cmatter}). Using this junction condition, one can rewrite the induced metric in the following form\footnote{Here, $g$ is extended with the region from $0$ to $2 g_c$. The region $g_c < g < 2g_c$ is the $\mathbb{Z}_2$-symmetric copy of the other region.}:
\begin{align}
ds^2 &= - \frac{f(z)}{z^2} dt^2 + \frac{1}{z^2(\sigma+\epsilon(z))^2} \frac{dg^2}{f(z)}\nonumber\\
&=-\frac{1-Z(g)^2}{z_h^2 Z(g)^2} dt^2 + \frac{L^2}{z_h^2 Z(g)^2 \left(\tilde{\sigma}+\tilde{\epsilon}(g)\right)^2}\frac{dg^2}{1-Z(g)^2}\,.
\end{align}
This metric has the following scalar curvature:
\begin{align}\label{Curvature2D}
R^{[h]}(g) =- 2 \frac{z_h^2}{L^2} \left(\tilde{\sigma}+\tilde{\epsilon}(g)\right)^2 \left(Z'(g)^2- Z(g)Z''(g) -\frac{Z(g)Z'(g)\tilde{\epsilon}'(g)}{\tilde{\sigma}+\tilde{\epsilon}(g)} \right)\,.
\end{align}

This geometry has two $AdS_2$-asymptotic boundaries. This can be seen easily by taking the $Z\to 0$ limit. From Figure \ref{fig:Phy}, one can see that the energy density vanishes at the boundary thus the curvature becomes negative as we expect. On the other hand, the curvature approaches the following form at the tip of the EOW brane:
\begin{align}
R^{[h]}(g_c) \to  2 \frac{z_h^2}{L^2} \left(\tilde{\sigma}+\tilde{\epsilon}_c\right)^2 \tilde{z}_c Z''(g_c)\,.
\end{align}
Since $Z''$ at the tip is negative, the curvature of the tip is also negative.

As we discussed earlier, the observer confined on the EOW brane only detects the metric or the curvature. Also, this observer may construct an effective model to describe its world. One possible model is the JT gravity conformally coupled to matters as follows:
\begin{align}
\mathcal{I}_{JT+matt.} = \frac{1}{2}\int d^2 x \sqrt{-h}\phi \left(R +\frac{2}{L^2(1+\tau^2)} + \mathcal{L}_{\text{matter}} \right)\,.
\end{align}
Or, another effective model can be the deformed JT gravity proposed in \cite{Witten:2020ert}. The action of the model is given by
\begin{align}
\mathcal{I}_{deform} = \frac{1}{2}\int d^2 x \sqrt{-h}\left(\,\phi R - W(\phi) \right) \,.
\end{align}
In this case, the dilaton potential $W(\phi)$ can be related to the curvature (\ref{Curvature2D}) directly. It would be interesting to find a deformed JT model which produces this transition.

In the bulk description, the system is governed by gravity, the brane with tension and some energy-momentum tensor living on the EOW brane. However, the observer on the brane only feels the curvature or the metric. The above JT models are just effective descriptions, thus it is notable that the dilaton has no meaning in the bulk.

\subsection{EOW inside the horizon}

\begin{figure}[h!]
    \centering
    \includegraphics[width=8cm]{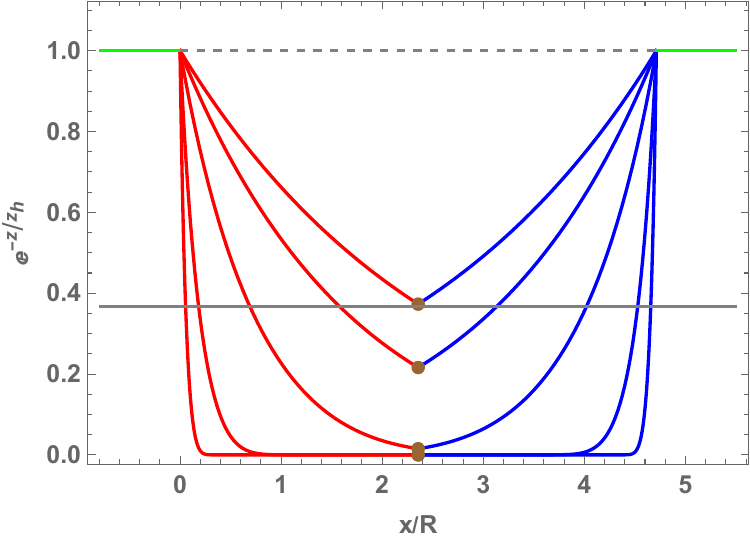}
    \includegraphics[width=8cm]{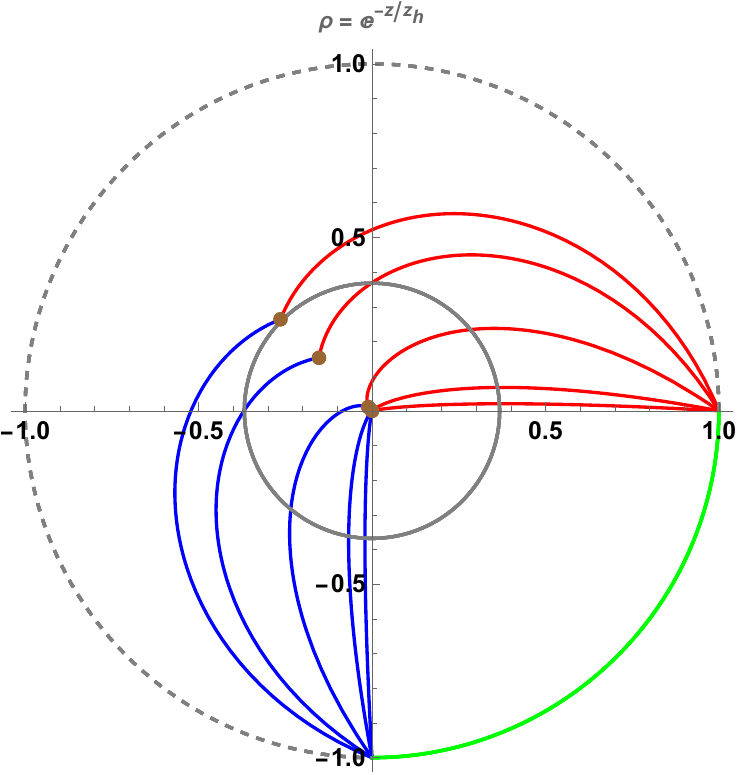}
    \caption{EOW brane configurations inside the horizon: The right figure is the same configuration as the left one in the polar coordinates $(\rho,\theta)$. The green line is the BCFT space, and the brown dots denote the intersecting point of two EOW branes. The gray solid line stands for the location of the horizon. As the temperature is lowered, the intersecting point moves to the horizon. The temperatures $T$ are given by $TR=1, 0.3, 0.08, 0.05, 0.035, 0.029$ and $0.0231$. This temperature-varying configurations are connected to Figure \ref{fig:EOWs01} continuously.}
    \label{fig:InEOW}
\end{figure}

In the previous subsection, we focus on the evolution under varying temperatures outside the horizon. However, it is possible to extend the EOW branes to the interior of the black hole. In this extension, the induced metrics have the interior regions with opposite signs of the time and space components like usual Lorentzian black hole metrics. We are still taking the symmetric case with $\sigma_1=\sigma_2=\sigma$. Let us look at the extended part of the EOW branes. These two branes touch each other inside of the horizon. By the similar reasoning for the outside of the horizon, we deduce that these two branes can be combined into one single brane even inside the horizon. We plot such configurations in Figure \ref{fig:InEOW}.

Figure \ref{fig:InEOW} shows that the intersecting point moves to the horizon by lowering the temperature. However, we are not sure how to describe the matter in this situation because of the opposite signs of the time and space components of the metric inside the horizon. Even though this interior physics is unclear, we believe that the intersecting point describes a scale governing interior physics.

In more detail, a merged single brane involving energy excitation like the outside configuration may describe a possible brane configuration of the interior. Then, the brane location is determined by a curve touching the horizon and connected to the outside branes. In addition, the energy excitation makes the brane detoured from the singularity of the BTZ black hole. Also, the excited energy configuration can't escape from the interior region. Therefore, the outside brane is governed by only the tension without matter. Whatever brane configuration inside the horizon can be formed, the length scale showing nontrivial physics is determined by the distance between the intersection point and the singularity position. Thus we will present the location of the intersecting point in terms of temperatures.

As one can see in Figure \ref{fig:InEOW}, the intersecting point seems to leave the origin suddenly at a specific temperature. However, this observation is not true since it just appears in our particular coordinate system with $\left(\rho\,, \theta\right)$. Thus we need to consider this observation in a different coordinate system.

 \begin{figure}[h!]
    \centering
    \includegraphics[width=7.5cm]{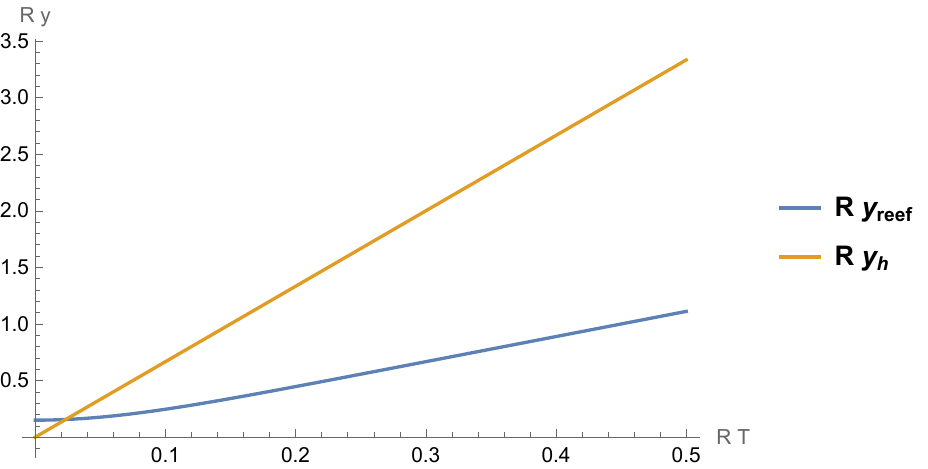}
    \includegraphics[width=7.5cm]{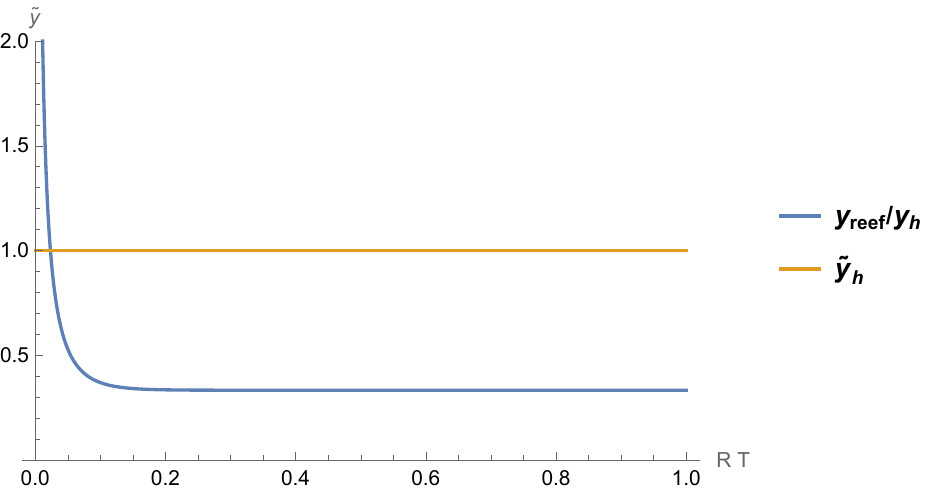}
    \caption{Comparison of the intersection point $y_{\text{reef}}$ and the horizon $y_h$: We set $\sigma L= 1/3$ and $L=1$, and the normalized radial coordinate is defined by $\tilde{y}=y/y_h$. }
    \label{fig:Reef}
\end{figure}

Let us look at the same phenomenon in terms of the coordinates $y$ and $\theta$, which appeared in the metric (\ref{metic2D}). By symmetry, the intersection point is determined by $L g = \frac{1}{2}\left(2\pi R - \mathcal{V}_1\right)$. Using this $y$-coordinate, the location of the intersection  point and the horizon are given as follows:
\begin{align}
&y_{\text{reef}} =\frac{\tau}{z_h}\coth \left(\frac{\pi R - \frac{\mathcal{V}_1}{2}}{z_h L}\right)=2\pi L T\tau \coth \left(\pi T\left(2\pi R - \mathcal{V}_1\right)\right)\,,\nonumber\\
&y_h =\frac{\sqrt{1+\tau^2}}{z_h}=2\pi L T\sqrt{1+ \tau^2}\,.
\end{align}
We plot these locations under the temperature variation in Figure \ref{fig:Reef}. We will call the region surrounded by the intersection point ``reef''. The singularity location of the BTZ black hole is commonly regarded as $y=0$. The EOW branes starting from the AdS boundary enter the interior of the horizon and touch each other before contacting the the singularity of BTZ black hole. Therefore, the induced metric of the EOW branes can not be extended to $y=0$. In the view of the EOW branes, the reef is formed and it hides the bulk singularity. The size of the reef $y_{\text{reef}}$ depends on the tension parameter. Also, the reef size approaches the horizon ($y_{\text{reef}}\to y_h$) as the tension increases.

As the figure shows, the horizon and the reef size are decreasing with the same ratio as lowering the temperature in the high-temperature region. At this region, the reef size becomes $2\pi L T \tau$. Interestingly, it can be related to the wrapping factor in the grafted thermodynamics (\ref{first law 00}). Therefore, we identify the reef size at high temperatures with the wrapping factor through the following relation:
\begin{align}
y^{\infty}\equiv\lim_{T\to\infty} y_{\text{reef}}=2\pi L T \tau=\frac{L}{4\pi G_{(3)}} \mathbf{w}\,.
\end{align}
This relation can be understood as follows. If we turn off the tension, then the wrapping factor vanishes. So the thermodynamics of JT gravity is decoupled from the thermodynamics of the BTZ black hole. In this tensionless limit, the EOW brane approaches the bulk singularity directly. Therefore, one may say that the existence of the reef makes it possible to join two different thermodynamics.

Now, let us consider the low-temperature region in Figure \ref{fig:Reef}. The reef behavior changes suddenly at low temperatures. The rescaled right figure shows this change more clearly. This behavior happens near the following temperature:
\begin{align}
T_{grow}= \frac{1}{\pi\left(2\pi R-\mathcal{V}_1\right)}\,.
\end{align}
The entropy of the JT black hole and the BTZ black hole at this temperature are given by
\begin{align}
S_{JT}^{grow}=\frac{4 \pi  L \sqrt{\tau ^2+1}}{2\pi R-\mathcal{V}_1} ~,~\mathcal{S}^{grow}=\frac{\mathcal{V}_1 L}{2 G_{(3)}\left(2\pi R-\mathcal{V}_1\right)}\,.
\end{align}
On the other hand, we already have another important scale. It is the temperature $T_{out}$ in (\ref{Temp-out}), where the horizon disappears. In terms of $\tau$, the temperature can be written as
\begin{align}
T_{out}=\frac{1}{2 \pi  (2 \pi  R-\mathcal{V}_1 )}\log \left(\frac{1+\frac{\tau }{\sqrt{\tau ^2+1}}}{1-\frac{\tau }{\sqrt{\tau ^2+1}}}\right)\,.
\end{align}
Using this expression, one can see that a temperature ratio $T_{out}/T_{grow}$ depends only on the tension. Therefore, a condition $T_{out}/T_{grow}<1$ restricts the range of tension parameter, i.e., $\tau<\tanh(1)$. In this parameter region, the reef can be formed inside the horizon. Otherwise, the two EOW branes meet outside the horizon.

Let us consider a special case with a constant wrapping factor. In this case, the tension parameter is inversely proportional to the temperature. The physical meaning of this case is that the high-temperature reef size $y^{\infty}$ is taken to be fixed. Then, the grafted thermodynamics is given as follows:
\begin{align}
\delta (E_B - 2\mathbf{w}  E_{JT})= T \delta (\mathcal{S}_{total}-2\mathbf{w}  S_{JT} ) - \mathcal{P} \delta \mathcal{V}_1\,.
\end{align}
This thermodynamic relation describes a thermodynamic system of the following effective energy and entropy with volume $\mathcal{V}_1$:
\begin{align}
E_{\text{eff}}= E_B -2 \mathbf{w}\,E_{JT}~,~S_{\text{eff}}= \mathcal{S}_{total} - 2\mathbf{w}\,S_{JT}\,.
\end{align}
These quantities can be negative for small $\mathcal{V}_1$. We will see when this negativity can happen. To see this, we take a high-temperature limit. Then these quantities become
\begin{align}
E_{\text{eff}}\sim \frac{\left( \mathcal{V}_1 - 16\pi G_{(3)}\mathbf{w}\right)L\pi}{4G_{(3)}} T~,~S_{\text{eff}}\sim \frac{\left( \mathcal{V}_1 - 16\pi G_{(3)}\mathbf{w}\right)L\pi}{4G_{(3)}} T^2\,.
\end{align}
Therefore, the size of the BCFT${}_2$ volume $\mathcal{V}_1$ should be larger than the reef size $\mathbf{w}=y^{\infty}$ with the three-dimensional Newton constant to have well-defined thermodynamics.

\subsection{Summary of EOW brane evolution by lowering temperature}

 \begin{figure}[h!]
    \centering
    \includegraphics[width=3.8cm]{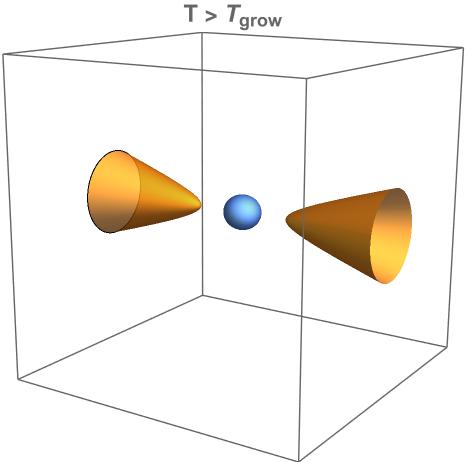}
    \includegraphics[width=4.2cm]{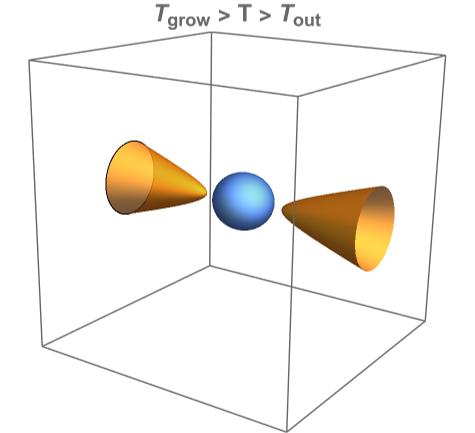}
    \includegraphics[width=3.8cm]{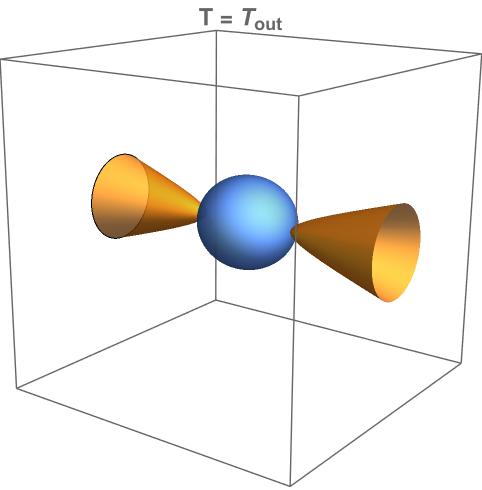}\\
    \includegraphics[width=3.8cm]{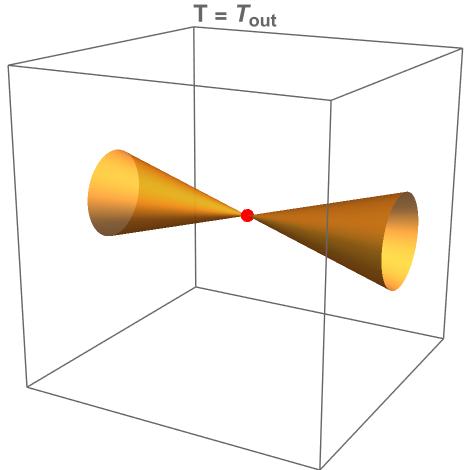}
    \includegraphics[width=3.8cm]{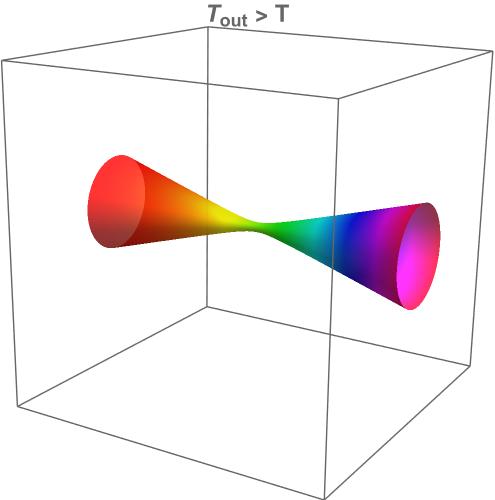}
    \includegraphics[width=3.8cm]{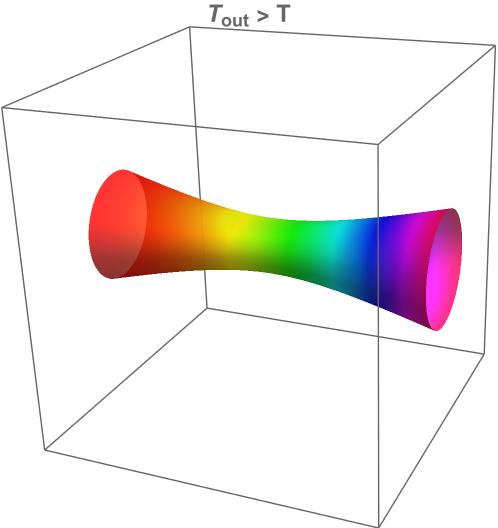}
    \caption{Cartoons from the observer point of view on the EOW branes}
    \label{fig:Cartoons}
\end{figure}

This subsection summarizes how the EOW brane evolves by lowering the temperature. We provide cartoons in Figure \ref{fig:Cartoons} to visualize the evolution.

We are aware of the full-bulk geometry (BTZ black hole) and the EOW branes. Let us assume that the observers on the boundary of the EOW branes can not see the bulk. These two observers recognize their horizons in high temperatures ($T>T_{grow}$). The brane configuration tells us that the reef region formed already in high temperatures. It surrounds the singularity of the BTZ black hole. However, the observers can not see this reef. When the temperature becomes $T_{grow}$, the reef starts growing relative to the horizon size. Lowering the temperature to $T_{out}$, the reef touches the horizon. Each observer sees each horizon disappear, and a very high (delta function-like) energy density appears at the previous horizon position. In low temperatures ($T<T_{out}$), the two spacetimes are connected, and the energy flux flows toward each observer. Also, two observers are now sitting on a single spacetime below this temperature. The spacetime is nothing but the asymptotically AdS space with two asymptotic boundaries.

Our description follows an observation of the EOW brane evolution under the temperature variation. We agree that many details are ignored in this scenario. We leave these details for future studies. Also, there is an important notice in this result. From the bulk point of view, The locations of EOW branes and their induced geometries are governed by the junction equation. However, the observers living on the EOW branes regard their geometries as the JT gravity systems. This is a notable lesson in this work.

%%%%%%%%%%%%%%%%%%%%%% 
\section{Discussion and Future Direction} 
%%%%%%%%%%%%%%%%%%%%%% 

In this work, we generalize the thermodynamic first law with the tension parameter for the BCFT${}_2$ system dual to the BTZ black hole with the EOW branes. And the tension contribution can be replaced with the first law of the JT black hole. So, we combine two first laws into a grafted thermodynamic law given in (\ref{first law 01}). The resultant law has a factor that respects the EOW brane structure inside the horizon. This wrapping factor is identified with the reef size in high temperatures higher than $T_{grow}$. From the point of view of the JT system, the thermodynamic relation (\ref{1st law JT}) includes the effective cosmological constant variation associated with the thermodynamic volume and pressure. It is interestingly notable that the PV criticality introduced in \cite{Dolan:2011xt, Kubiznak:2012wp, Kubiznak:2016qmn} can be realized in the JT system. We leave this as a future direction. The form of the thermodynamic first law can have a universal formulation even in higher dimensions. We provide a comment with an expected expression (\ref{first law B}) in Appendix B. However, this universal form induces a nontrivial matter excitation on the EOW branes. Such a matter configuration is another issue that will be addressed later.

We provide a scenario summarized in Figure \ref{fig:Cartoons} for the two JT black hole systems. There are two significant temperature scales $T_{grow}$ and $T_{out}$ governing the interior structure of the JT black holes. We introduce the reefs that are intimately related to these temperatures. This reef looks so similar to the island in the Page Curve study. However, we know that the two concepts are different by definition. More studies should be accomplished to reveal whether they are related to each other or not. We will also study this question as a future problem. It is another important subject to explore what kind of physics corresponds to this transition in the BCFT description. Unfortunately, it is beyond the scope of the present work. It could be related to the boundary degrees of freedom of the BCFT under the temperature variation. We hope to present such a study soon.

Also, it would be intriguing to extend our system to higher dimensions. We expect that there will be richer structures in this generalization. We may generalize the bulk geometry with charges or angular momentum, and then one can ask what kind of black hole can exist on the EOW branes. In addition, this nontrivial structure on the EOW brane should have a holographic interpretation at the boundary of the BCFT. For instance, the edge state may be related to the EOW brane structure in this double holography picture. We leave this application as another future direction.

Notably, the induced metric does not seem to respect the physical matter on the EOW branes from the bulk point of view. Since, however, this matter respects the holographic g-theorem introduced in \cite{Takayanagi:2011zk}, the energy conditions of the matter are somehow related to the EOW configuration. Anyway, the JT black hole system is an excellent model to figure out the EOW observers' world. The tension of the bulk gravity only appears as an effective cosmological constant in the braneworld. This EOW perspective must be more critical in higher dimensions. In such cases, the EOW brane can have black hole metrics. We provide the result partly in Appendix B. On the other hand, the corresponding matter on the brane could have a nontrivial configuration. The energy condition of this matter could be an important issue.

As a final comment, our transition from the black hole to the horizonless geometry appears at the temperature $T_{out}$. Also, it is well known there is a phase transition between the BTZ black hole and the thermal AdS (or AdS soliton) in the BCFT setup \cite{Fujita:2011fp, Takayanagi:2011zk} like a confinement phase transition. Our transition can be compared to this first-order phase transition. If the BCFT volume is quite large, $T_{out}$ becomes very high. Therefore, the validity of our description depends on how the BCFT volume is large enough. It would be interesting to accompany the first-order phase transition with our construction. We will report the analysis soon.

\section*{Appdendix}

\section*{A. Shadow Entropy = Boundary Entropy in 2D}

\begin{figure}[h!]
    \centering
    \includegraphics[width=11cm]{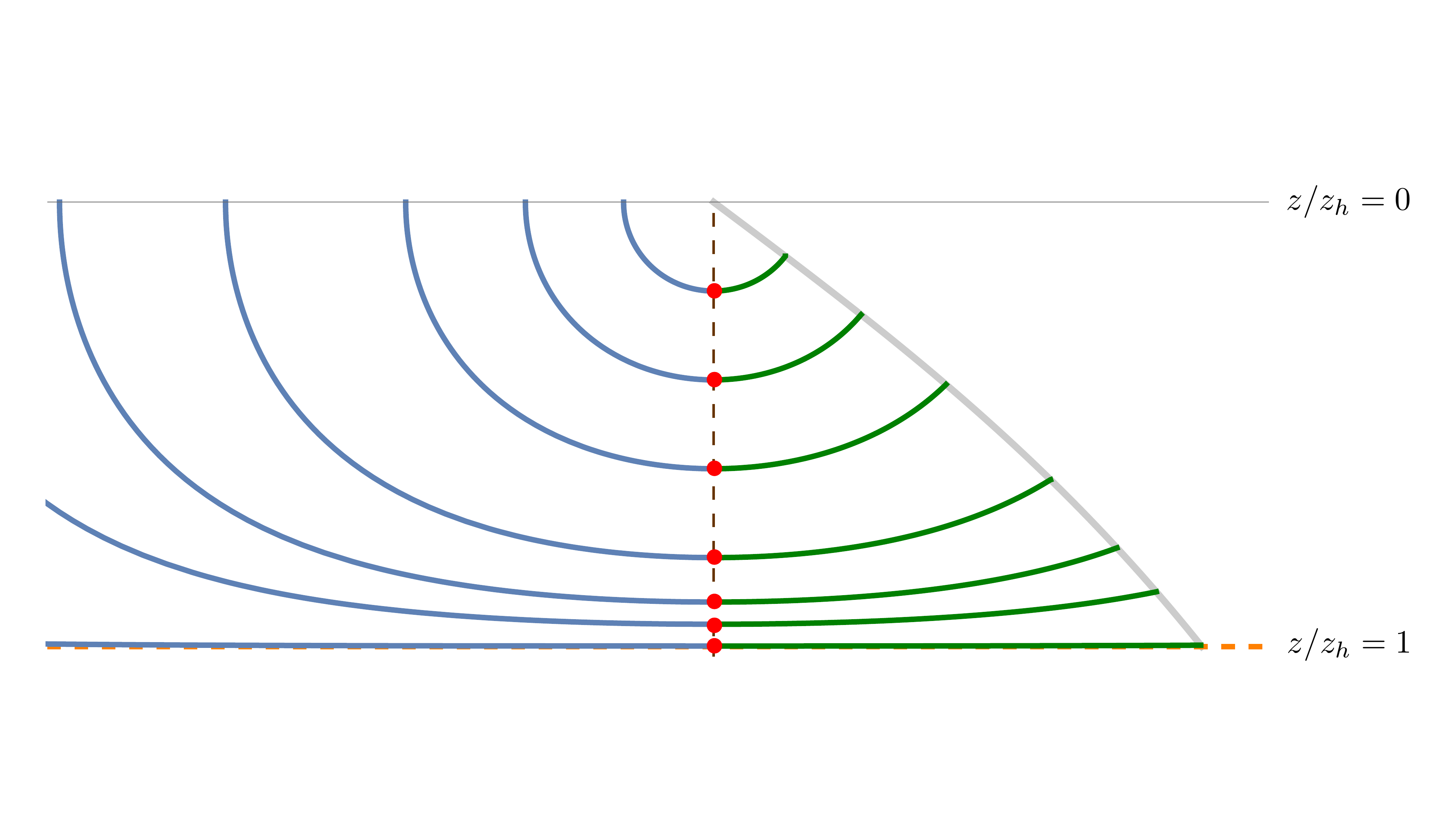}
    \caption{Ryu-Takayanagi surfaces (Blue and green lines) that intersect with the EOW brane. The intersection point is denoted by $(g_Q,z_Q)$. Red dots indicate the IR turning point of each minimal surface: from top and bottom, $z_0=0.2, 0.4, 0.6, 0.8, 0.9, 0.95, 0.999$. Here we choose $L=1$, $\sigma=0.8$ and $z_h=1$.}
    \label{fig:RT}
\end{figure}

This section shows the equivalence between the shadow entropy and the boundary entropy. The boundary entropy can be obtained by computing the on-shell action in \cite{Takayanagi:2011zk}. It turns out that this boundary entropy is the same as the length of minimal surfaces denoted by the green lines in Figure \ref{fig:RT}.

%%%%%%%%%%%%%%% Figure %%%%%%%%%%%%%%%%%%%%%%%%%%%%%%%%%
%%%%%%%%%%%%%%%%%%%%%%%%%%%%%%%%%%%%%%%%%%%%%%%%%%%%

Let us compute the entanglement entropy by utilizing Ryu-Takayanagi's formula
\begin{align}
S_{\text{EE}}=\frac{{\cal{A}}(\Sigma_{min})}{4G_N},
\end{align}
where $\Sigma_{min}$ is the codimension two extremal bulk geodesic which is homologous to the subregion of the conformal boundary and 
${\cal{A}}(\Sigma_{min})$ is the area of the minimal surface. The center of the subregion is located at a boundary point of the BCFT${}_2$. For $g(z)>0$, then the intersection point of the EOW brane and the Ryu-Takayanagi's surface is denoted by $(g_Q, z_Q)$. This point can be found in terms of the turning point $z_0$ of the minimal surface.
\begin{align}
g_Q(z_0)=&z_h \cdot \sinh^{-1}\biggl(\frac{L\sigma z_0}{\sqrt{z_h^2-L^2\sigma^2z_0^2}}\biggr), 
\label{intergq}
\\
z_Q(z_0)=&\frac{z_h z_0\sqrt{1-L^2\sigma^2}}{\sqrt{z_h^2-z_0^2L^2\sigma^2}}.
\label{interzq}
\end{align}
The length of the minimal surface between the intersecting point and the tip $z_0$ of the minimal surface can be computed as follows:
\begin{align}\label{AAA}
{\cal{A}}(\Sigma_2)=L\int_{z_Q(z_0)}^{z_0}\frac{z_0z_h}{z\sqrt{(z_h^2-z^2)(z_0^2-z^2)}}dz.
\end{align}
This expression is available for all the green lines in Figure \ref{fig:RT}. In addition, this length is independent of $z_0$ and $z_h$. One can show  $\frac{\delta \mathcal{A}}{\delta z_0}=\frac{\delta \mathcal{A}}{\delta z_h}=0$ using the above expression. This property enables us to take the limit $z_0\to z_h$. Then $z_Q$ also approaches to $z_h$. This means that $\mathcal{A}$ becomes the shadow entropy from the boundary point. This shadow entropy is given by
\begin{align}\label{shadow entropy}
S^{(i)}_{\text{\tiny Shadow}}=\frac{1}{4G_N}\int^{g_i} dx \,g_{xx}=\frac{L}{4G_N z_h}\int^{g_i} dx=\frac{c}{6}\cdot \tanh^{-1} (L\sigma_i),
\end{align}
where the Brown-Henneaux relation is used to express the gravity parameters $L$ and $G_N$ in terms of the central charge, $c=\frac{3L}{2G_N}$. This shadow entropy is the same as the boundary entropy computed in \cite{Takayanagi:2011zk}. However, this equivalence does not hold in higher dimensions in a naive manner of extension.

\section*{B. Comment on Grafted Thermodynamics in Higher Dimensions}

Now, we extend the previous EOW setup to higher dimensions. We focus on the Schwarzschild black brane case. The metric of black brane can be written as
\begin{align}
ds^2 = \frac{1}{z^2} \left( -f(z) dt^2 + \frac{L^2}{f(z)}dz^2 + \sum_{i=1}^{d-1}dx^i dx^i\right)\,.
\end{align}
We introduce two $(d-1)$-dimensional EOW branes in this bulk. The location of an EOW brane is determined by $x^{d-1}= L g(z)$. To obtain the extrinsic curvature in a convenient coordinate system. We introduce a coordinate $y$ satisfying $y = L g(z)-x^{d-1}$. Then, one can find the metric in terms of $y$ as follows:
\begin{align}\label{bulk D metric}
ds^2 =& \frac{1}{z^2}\left(-f dt^2 + \sum_{i=1}^{d-2} dx^i dx^i \right) +\frac{L^2}{z^2}\frac{1+g'^2 f}{f} \left( dz -\frac{g' f}{L(1+g'^2 f)}dy  \right)^2\nonumber\\
&+ \frac{dy^2}{z^2\left(1+ g'^2 f\right)}\,.
\end{align}
Under the ADM decomposition, the laps function and the nontrivial component of the shift vector are given by
\begin{align}
N =\frac{1}{z\sqrt{1+g'^2 f}}~,~\mathcal{N}^z = -\frac{f g'}{L\left(1+g'^2 f\right)}\,.
\end{align}
Also, the induced metric can be read off from the bulk metric (\ref{bulk D metric}). It is
\begin{align}\label{met_ind}
ds_{ind}^2 = \frac{1}{z^2}\left(-f dt^2 + \sum_{i=1}^{d-2} dx^i dx^i \right) + \frac{L^2}{z^2}\frac{1+g'^2 f}{f} dz^2\,.
\end{align}
Using the above expressions, one can construct the junction equation (\ref{junctionEq}). However, there is some difference from the previous $d=2$ case. We will explain this step in detail below.

The junction equation is 
\begin{align}
K_{ab}- K h_{ab} - \sigma h_{ab} = T_{ab}\,.
\end{align}
In the two-dimensional case, the energy-momentum tensor on the EOW brane $T_{ab}$ vanishes when the induced metric has a constant curvature. However, it is not guaranteed in higher dimensions. Thus we consider firstly an induced metric with constant curvature. To do this using the induced metric, we find the curvature expression in general dimensions, which is given by
\begin{align}
R[h]=\frac{z^{d+1}}{2 L^2} \frac{\partial }{\partial z}\left(\frac{d+ (d-2) f}{z^{d} \left(1+f {g'}^2\right)}\right)\,.
%\frac{d \left\{d-(d-2) f^2-2 d\, f\right\}g'^2-2 d (d-1)-2 f z \left\{d+(d-2) f\right\}g' g''}{2 L^2 (1 + f g'^2)^2}\,.
\end{align}
Here we used the background Einstein equation, $f'(z)= -\frac{d}{z}(1-f(z))$. We are devoted to the construction of local AdS spaces in this work. Thus, we give a constraint for constant curvature, $R[h]= - d(d-1)/L^2_{\text{eff}}$. This condition and the regular infalling EOW brane at the horizon lead to the following differential equation for $g(z)$:
\begin{align}\label{dg general D}
g'(z)=\sqrt{\frac{2(d-1) \left(\ell_{\text{eff}} ^2-1\right)}{\left(2\, (d-1)-d\, \ell_{\text{eff}} ^2\right) f(z)+d\, \ell_{\text{eff}} ^2}}\,,
\end{align}
where $\ell_{\text{eff}}$ is defined by a ratio $\ell_{\text{eff}}=L_{\text{eff}}/L$. The location of the EOW brane can be solved as the following function:
\begin{align}
g(z)-g(0) = \sqrt{\ell_{\text{eff}} ^2-1}\,z   \,\, _2F_1\left(\frac{1}{2},\frac{1}{d};\frac{1}{d}+1;-\frac{\left(d \left(\ell_{\text{eff}}^2-2\right)+2\right)}{2 (d-1)}\left(\frac{z}{z_h}\right)^d\right)\,.
\end{align}
This EOW brane configuration guarantees that the induced metric (\ref{met_ind}) has constant curvature. Also, one can notice that there exists a horizon and the temperature is the same as the temperature of the bulk black brane by the form of the induce metric (\ref{met_ind}). Near the conformal boundary $z=0$, the EOW brane becomes a straight line. Therefore, the tension parameter is related to the effective AdS radius $L_{\text{eff}}$.

Now, let us comment on the matter on the EOW brane. In higher dimensions with the constant curvature EOW brane, the matter living on the EOW brane does not vanish. This is the main difference from the $d=2$ case. There is the tension matter together with the nontrivial matter excitation which gives a constant scalar curvature $R[h]$.

To construct the grafted thermodynamics in higher dimensions, the total entropy $\mathcal{S}_{\text{total}}$ plays a crucial role. This entropy is again identified with the total horizon area. In higher dimensions, however, the increased area by the EOW brane depends on the temperature. More explicitly, the shadow entropy of i-th brane takes the following form:
\begin{align}
S^{(i)}_{shadow}=\frac{1}{4 G_{d+1}}\frac{\mathcal{V}_{d-2}}{z_h^{d-2}}\mathbb{G}(L_{\text{eff}}(\sigma^i))\,,
\end{align} 
where $\mathbb{G}$ does not depends on $z_h$. The factor $z_h^{d-2}$ in the denominator gives nontrivial contribution to the entropy variation $\delta S_{shadow}^{(i)}$. So the internal energy is not directly identified with the black hole mass of the $AdS_{d+1}$ black brane due to this temperature dependence. Therefore, we need to define a new internal energy $U$. Using this internal energy, we may speculate that the first law of the grafted thermodynamics in higher dimensions is as follows:
\begin{align}\label{first law B}
\delta U - T \delta\mathcal{S}_{total}  +\mathcal{P}  \delta\mathcal{V}_{d-1} = \sum_i^2 \mathbf{w}_i   \left(  \delta E^i_{EOW}- T \delta S^i_{EOW} \right)\,.
\end{align}  
We hope to present more details on this expression soon.

\section*{Acknowledgments}
We thank Sang-Jin Sin, Xian-Hui Ge and Byoungjoon Ahn for their helpful discussions.
This work is supported by Basic Science Research Program through NRF grant No. NRF-2022R1A2C1010756(Y.Seo), NRF-2019R1A2C1007396(K.K.Kim). C. Park was supported by the National Research Foundation of Korea (No. NRF-2019R1A2C1006639). J.
H. Lee was supported by the National Research Foundation of Korea(NRF) grant funded
by the Korea government(MSIT) (No. NRF-2021R1C1C2008737). K.K.Kim acknowledges
the hospitality at APCTP where part of this work was done.

%%%%%%%%%%%%%%%%%%%%%%%%%%%%%%%%%%%%%%%%%%%%%%

\end{document}